\newif\ifarxiv
\arxivtrue   

\documentclass[]{jfm}

\usepackage{graphicx}
\usepackage{newtxtext}
\usepackage{newtxmath}
\usepackage{natbib}
\usepackage{hyperref}
\usepackage{gensymb}
\usepackage{wrapfig}
\hypersetup{
    colorlinks = true,
    urlcolor   = blue,
    citecolor  = black,
}

\newcommand{\RomanNumeralCaps}[1]
\linenumbers

\ifarxiv
\makeatletter

\let\ps@titlepage\ps@plain

\def\@maketitle#1{%
  \newpage
  \thispagestyle{plain}%
  \vspace*{2em}
  {\centering
    {\LARGE\bfseries \@title \par}
    \vskip 1.5em
    {\large \@author \par}
    \vskip 1em
    {\small \@affiliation \par}
  }
  \vskip 2em
}

\makeatother
\fi

\ifarxiv
\makeatletter
\renewcommand{\email}[1]{}
\makeatother
\fi

\ifarxiv
\makeatletter
\renewcommand{\pagelimitfooter}{}
\makeatother
\fi

\title{On the role of grazing atmospheric particles in hypersonic boundary-layer receptivity}

\author{Sean Dungan\aff{1},
 \and Christoph Brehm\aff{1}}

\affiliation{\aff{1}Department of Aerospace Engineering, University of Maryland, College Park, MD 20742, USA}

\begin{document}

\maketitle

\begin{abstract}

 The role of small atmospheric particulates in the initiation of the laminar-turbulent transition is investigated for the HIFLIER vehicle during Mach 6 ascent condition. High-fidelity simulations are used to examine the disturbance wavepackets generated by particles following either grazing trajectories through the boundary layer or trajectories that lead to impact and rebound at the cone surface. Grazing particles that penetrate less than halfway through the boundary layer are found to generate wavepackets with amplitudes comparable to those produced by wall-impinging particles. To identify the receptivity mechanisms responsible for this behaviour, a multiple-scales asymptotic formulation is developed that accounts for the extended streamwise region over which a particle interacts with the boundary layer. The analysis shows that a grazing particle can generate a strong disturbance response when crossing the critical layer, where the streamwise-momentum component of the adjoint mode is largest and the boundary layer is most receptive. Thus, direct surface impact is not required for atmospheric particulates to excite dynamically significant instability waves. Root-mean-square disturbance estimates for submicron grazing particles predict transition-onset locations close to those measured in flight, suggesting they may have contributed significantly to the onset of transition on HIFLIER at the trajectory condition considered.
\end{abstract}

\section{Introduction}

Laminar-turbulent transition (LTT) for high-speed vehicles has several persistent challenges to its full understanding and, therefore, prediction \citep{Chou2024}. It appears desirable to move towards the amplitude method \citep{Mack1977}, but doing so requires a receptivity solution for the initial disturbance amplitudes specific to a disturbance source. While acoustics provide the dominant disturbance source initiating transition in wind tunnels \citep{schneider2008}, the source(s) in free flight have only been postulated \citep{bushnell1990}. The most likely candidates for causing transition in hypersonic free flight include kinetic fluctuations and atmospheric aerosols, which have measurements \citep{habeck2020}.

\cite{Fedorov2013} provided the first theoretical analysis of boundary-layer receptivity to disturbances generated by particulates in supersonic flow. The model was formulated within a temporal stability framework and  assumed that the particle–boundary-layer interaction occurred over a sufficiently short streamwise distance for the locally parallel approximation to remain valid. Subsequent studies have employed full Navier–Stokes simulations to investigate particulate receptivity over a range of geometries and flow conditions. Good agreement was achieved for all the available numerical solutions \citep{browne_particle_2021,Habeck2026}, but the theoretical predictions of particle-excited first mode amplitudes differed from the numerical solutions by up to 40\%. Efforts have begun very recently to capture the particle-induced transition scenario in a statistical model like \citet{al_hasnine_particle_model_2023}. \cite{Chuvakhov2025} used the receptivity solutions from \cite{Fedorov2013} to model LTT onset for particles distributed in radius. They found micron-sized particles contributed most to the predicted LTT location, but in some instances submicron particles were also relevant to the calculation. While these tiny particles are able to reach the surface of the $14\degree$ sharp wedge considered there, they tend to only graze the boundary layer as the vehicle geometry becomes more slender and are unable to reach the surface. \cite{Habeck2026} showed through 2D numerical simulation how some of these `grazing' particle trajectories excite second-mode wavepackets of comparable strength as those impinging the vehicle's surface. Few details regarding why this occurs (i.e., the associated receptivity process) were provided, however.

\begin{figure}
    \centering
    \def\svgwidth{0.9\textwidth}
    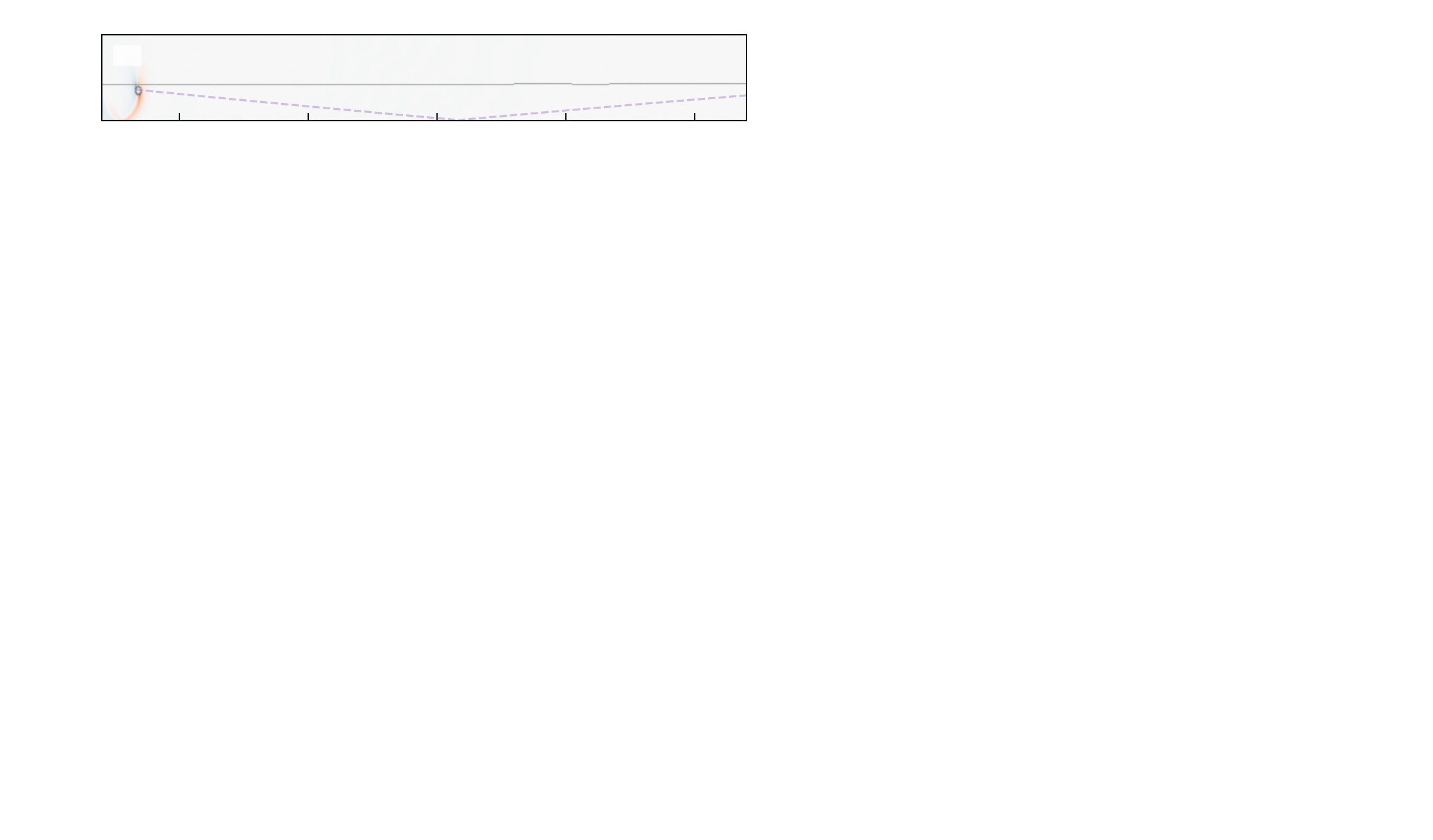
    \caption{Pressure disturbance field comparison between impinging (left) and grazing (right) trajectory scenarios with $\rho_{p}^{*}=0.50$ g/cm$^{3}$ and $\rho_{p}^{*}=0.25$ g/cm$^{3}$, respectively.}
    \label{fig:flow_field}
\end{figure}

Figure~\ref{fig:flow_field} highlights that a grazing particle can generate a wavepacket of comparable amplitude to that produced by an impinging particle, despite generating weaker acoustic radiation and fewer small-scale flow structures. The importance of grazing trajectories was also demonstrated in the statistical model developed by the present authors in \cite{al_hasnine_particle_model_2023}, where their inclusion increased the mean flux of particles capable of exciting wavepackets by approximately 185\%. 
Since the particle flux is a key input to statistical models, this study seeks to clarify the role grazing particles play in initiating LTT in hypersonic free flight. We first develop the method of multiples scales (MMS) receptivity formulation within a spatial stability framework to model the wavepackets excited by grazing particulates over extended streamwise distance. After verifying the formulation against the numerical method of \cite{browne_particle_2021}, we use the MMS analysis to elucidate the receptivity mechanisms associated with both impinging and grazing particle trajectories. Finally, the statistical model of \cite{Chuvakhov2025} is used to show that the transition location predicted by grazing particles agree more closely with recent flight data (\cite{Borg2025}).

\section{Methodology} \label{sec:methodology}

\subsection{Flow conditions, geometry, and base flow} \label{sec:flow_conds}

The significance of grazing particles is examined using the geometry and flight conditions of the recent HIFLIER hypersonic flight experiment \citep{Borg2025}. We consider the freestream conditions at 26.95 seconds into ascent, when the vehicle is at an altitude of approximately $h_{\infty}=21.5$ km. Based on the data reported by \citet{Borg2025} and 1993 ICAO standard atmosphere, the prescribed freestream conditions are $P_{\infty}=4.359$ kPa, $T_{\infty}=218.1$ K, $M_{\infty}=6.25$, and $\Rey_{\infty}=9.05(10^6)$ 1/m. The geometry consisted of a slightly blunted straight cone with $7^{\circ}$ half angle, $2.5$ mm nose radius, and total length of $1.6$ m. 

An isothermal wall condition of $T_{w}=347$ K is imposed to compute a laminar base flow using the CHAMPS CFD framework utilising a body-fitted grid, 5th-order WENO for convective terms, and 2nd-order central discretisation for the viscous terms \citep{Mcquaid2024}. Grid convergence was verified by changing the initial wall spacing from $\Delta \eta_{w}=4$ $\mu$m to $1$ $\mu$m. The resulting change in surface heat flux near the outlet was less than 0.025\%, indicating that the laminar base flow solution is effectively grid independent.

\subsection{Solving the Disturbance Flow Equations} \label{sec:nlde_sec}

The numerical framework developed by \citet{browne_particle_2021} for simulating particle-induced LTT is employed within the in-house CHAMPS CFD solver. The method solves the Navier-Stokes equations in nonlinear disturbance equation (NLDE) form and uses an efficient adaptive-mesh-refinement, wavepacket tracking (AMR-WPT) algorithm. The present simulations use the particle-source-in-cell formulation, in which the particulate-induced momentum and energy source terms (see \citet{browne_particle_2021}) are added to the right-hand-side of the NLDE through a spatial kernel, $\delta^{*}(\mathbf{x}^{*}-\mathbf{x}_{p}^{*}(t^{*}))$, defined as
\begin{equation} \label{eq:kernels}
    \delta_{\mathrm{3D}}^{*} = \frac{1}{(\sigma^* \sqrt{2\pi})^{3}} \exp{\left(\frac{-|\mathbf{x}^{*}-\mathbf{x}_{p}^{*}(t^{*})|^2}{2 (\sigma^*)^2}\right)}, \quad \delta_{\mathrm{2D}}^{*} = \frac{1}{2\pi(\sigma^*)^2} \exp{\left(\frac{-|\mathbf{x}^{*}-\mathbf{x}_{p}^{*}(t^{*})|^2}{2 (\sigma^*)^2}\right)},
\end{equation}
where $()^*$ denotes dimensional quantities. The 2D kernel chosen this way directly yields the spanwise spectral density flow field for spanwise wavenumber $\beta^*=0$. This equivalence is demonstrated for the planar, second Mack mode in section \ref{subsec:traj_nlde}.

The particle position, $\mathbf{x}_{p}^{*}=\mathbf{x}_{p}^{*}(t^{*})$, in eq. \ref{eq:kernels} is obtained by integrating the Lagrangian equations of motion from prescribed initial conditions located immediately downstream of the conical shock. The drag coefficient for a spherical particle of radius $R_{p}^*$ is evaluated using the correlation of \citet{Singh2022}, which is applicable from the continuum to the free-molecular regime. The adaptive Cartesian mesh used to resolve both the particle source kernel and the resulting wavepacket had minimum cell spacings of $(\Delta x^{*}_{\mathrm{min}}, \Delta y^{*}_{\mathrm{min}})=(11.7, 6.29)$ $\mu$m for the two-dimensional simulations. The kernel width is specified as $\sigma^*/R_{p}^*=2.5$. Previous work has shown this choice accurately captures the particle-disturbance interaction while rendering the results insensitive to $\sigma^*$. We further verified this behaviour using simulations with $\sigma^*/R_{p}^*=5.0$, which produced a maximum difference 4.5\%  in downstream wavepacket amplitude. Only aerodynamic drag is included in the particle equations of motion. Local thermal equilibrium between the particle and the surrounding fluid is assumed, such that $T_p=T$ and the particle heat-source term vanishes, i.e., $\Bar{Q}_{p}=2\pi Nu (T_{p}-T)\equiv0$. 

\subsection{Method of Multiple Scales Particle Receptivity Model} \label{sec:mms_prtcl_model}

A method of multiple scales (MMS) formulation is developed to solve the boundary-layer receptivity problem subject to distributed forcing. A related formulation was derived by \citet{Edwards2019} for receptivity of kinetic fluctuations. The asymptotic analysis assumes that the base flow and the eigenfunctions of the locally parallel, homogeneous linear stability problem vary slowly in the streamwise direction. Accordingly, a slow streamwise coordinate is introduced as $x_{1}=\epsilon x$ with $\epsilon=1/R$ where the local Reynolds number is defined using boundary-layer-edge quantities as $R=\frac{U_{e}^{*}\Delta^{*}}{\nu_{e}^{*}}=\sqrt{\Rey_{x}}$. Here, $U_e^*$, $\Delta^*$ and $\nu_e^*$ denote the boundary-layer-edge velocity, characteristic length scale and kinematic viscosity. 

Substituting a WKBJ ansatz for the disturbance vector, together with the slowly varying (in $x$) base flow into the forced linearised Navier-Stokes equations, and applying the solvability condition yields an ODE for  the amplitude correction to the leading-order stability solution
\begin{equation} \label{eq:amp_ode}
    \frac{d c}{d x_{1}} + c W + \epsilon^{-1}q K^{-1}=0,
\end{equation}
where $W= \Gamma^{-1} \left\langle \mathbf{H}_{2}\frac{\partial \mathbf{A}_{0}}{\partial x_{1}}+\mathbf{H}_{3}\mathbf{A}_{0}, \mathbf{B}_{0} \right\rangle$ represents the non-parallel terms, $q=\Gamma^{-1}\left\langle \hat{\mathbf{G}},\mathbf{B}_{0} \right\rangle$ contains the Fourier transformed (in time and span) particle source terms $\hat{\mathbf{G}}$, and $K(x_{1})=\exp{\left[i\epsilon^{-1}\int_{x_{1u}}^{x_{1}}\alpha(\eta)d\eta\right]}$ is the WKBJ `phase.' The normalisation factor is defined as $\Gamma=\left\langle \mathbf{H}_{2}\mathbf{A}_{0},\mathbf{B}_{0}\right\rangle$ where the solutions of the direct and adjoint $\mathcal{O}(1)$ stability problems for a particular mode with eigenvalue $\alpha\in\mathbb{C}$ are $\mathbf{A}_{0}$ and $\mathbf{B}_{0}$, respectively. The definitions of matrices $\mathbf{H}_{2}$ and $\mathbf{H}_{3}$ are provided in \citet{Nayfeh1980}. Because the present study focuses on the excitation and subsequent exponential amplification of Mack's second mode, the non-parallel correction is neglected.  Equation \ref{eq:amp_ode} is integrated from an upstream location $x_{1,u}$ at which the particle forcing is negligible, i.e., $q\approx0$. With the direct eigenfunctions $\mathbf{A}_{0}$ normalised such that the flow quantity of interest has unity at the wall-normal location $y_i$, its dimensionless physical amplitude spectral density at the downstream location $x_{1}$ is given by
\begin{equation} \label{eq:mms_asd_eq}
    A(x_{1})=r_{p}^{2}c(x_{1})\mathbf{A}_{0}(x_{1},y_{i})K(x_{1}),
\end{equation}
where $r_{p}=R_{p}^{*}/\Delta^{*}$. Finally, the wavepacket envelope is obtained from the magnitude of the analytic signal of the disturbance quantity of interest. For example, for the wall-pressure fluctuation, the envelope is given by $\lvert p_a(x,z,t)\rvert$. Its maximum is estimated to leading order by applying the method of steepest descent about the saddle point $(\omega_{s},\beta_{s})$, yielding the following expressions for two- and three-dimensional wavepackets:
\begin{equation} \label{eq:2d_3d_mosd}
|p_a(x, z_s, t_s)| \sim \left( \frac{2^{1-j}}{\pi} \right)^{\frac{1-j}{2}} e^{N(x_d, x)} \left| \frac{A(x_{d})}{\left( S_{\omega\omega} S_{\beta\beta}^j - j S_{\omega\beta}^2 \right)^{1/2}} \right|_{(\omega_s, \beta_s)}
\begin{cases} 
j = 0, & \text{2D} \\ 
j = 1, & \text{3D} 
\end{cases},
\end{equation}
where $z_{s}$ and $t_{s}$ are the spanwise location and moment in time the envelope maximum associated with the saddle point wave of maximum $N$-factor, $N(x_{d},x)=-\Imag[S(x_{d},x)]$, $S(x_{d},x)=\int_{x_{d}}^{x}\alpha(\eta;\omega,\beta)d\eta$, passes over the given downstream observation location, $x$ (see \cite{Chuvakhov2025} for more details). The lower bound of integration, $x_{d}$, is some downstream location where the receptivity process is complete and the amplitude develops only due to exponential growth. The $A(x_{d})$ in eq. \ref{eq:2d_3d_mosd} is evaluated with eq. \ref{eq:mms_asd_eq} at that location.

The MMS particle receptivity model is implemented using the LST code from \citet{Tumin2007} to solve the direct and adjoint stability problems. Therefore, we take the inner product without complex conjugation, i.e., $\langle \mathbf{a},\mathbf{b} \rangle\equiv\int_{0}^{\infty} \sum_{i=1}^{8} a_{i} b_{i} dy$, making $\mathbf{B}_{0}$ the eigenfunction of the complex conjugate of the conventional adjoint stability problem. Since the time signal of the particle source terms is given implicitly through the spatial kernel and its trajectory, for each $(x,y)$ we generate a time signal and take the Fourier transforms for use in eq. \ref{eq:mms_asd_eq}. We normalise the eigenfunction to have unit pressure fluctuation at the wall for ease of comparison with NLDE simulation data. Finally, $x_{1,u}$ was chosen to be the position where the particle first reaches a wall-normal distance of $20\Delta^*$ and $q$ is effectively null.

\subsection{Grazing Study}

With all other parameters held fixed, the particle density, $\rho_{p}$, and initial (post-shock) position, $\mathbf{x}_{p,0}$, are varied to generate two classes of trajectories: impinging trajectories, for which the particle strikes and reflects elastically from the cone surface, and grazing trajectories, for which the particle penetrates the boundary-layer edge—defined by 99.5\% of the freestream stagnation enthalpy --- but does not contact the wall. To ensure a consistent comparison, the initial particle positions are selected such that all trajectories cross the boundary layer edge at the same streamwise location, $x^{*}=0.6$ m. This construction isolates the influence of the particle trajectory on modal excitation, since the particle source terms do not depend explicitly on $\rho_{p}$. The particle parameters considered in this study are summarised in table \ref{tab:grazing_study}.

\begin{table}
  \begin{center}
  \begin{tabular}{lccc}
    Trajectory Type & Impinging & Grazing & Grazing \\
    $\rho_p^{*}$ (g/cm$^3$) & $0.50$ & $0.25$ & $0.20$ \\
    $\mathbf{x}_{p,0}^{*}$ (m) & $(0.2790, \, 0.0259)$ & $(0.2090, \, 0.0208)$ & $(0.1790, \, 0.0188)$
  \end{tabular}
  \caption{Grazing study particle densities, initial positions and trajectory type.}
  \label{tab:grazing_study}
  \end{center}
\end{table}
These particle densities are set below those of representative atmospheric aerosols so that some particles with the selected radius, $R_{p}^{*}=5$ $\mu$m, follow grazing trajectories. These cases are intended to emulate the trajectories of submicron particles while retaining the same particle cross-sectional area, $A_p$, thereby enabling direct comparison of the disturbance amplitudes generated by grazing and impinging trajectories. This is appropriate because the particle-induced momentum force scales as $f_m \propto q_{\mathrm{rel}} A_{p} C_D$, where $q_{\mathrm{rel}}$ is the dynamic pressure based on the particle–fluid relative velocity,  and $C_D$ is the drag coefficient.

\section{Results}

\subsection{Trajectory and NLDE Simulations} \label{subsec:traj_nlde}

\begin{figure}
    \centering
    
    \begin{minipage}{0.48\textwidth}
        \centering
        \def\svgwidth{\linewidth}
        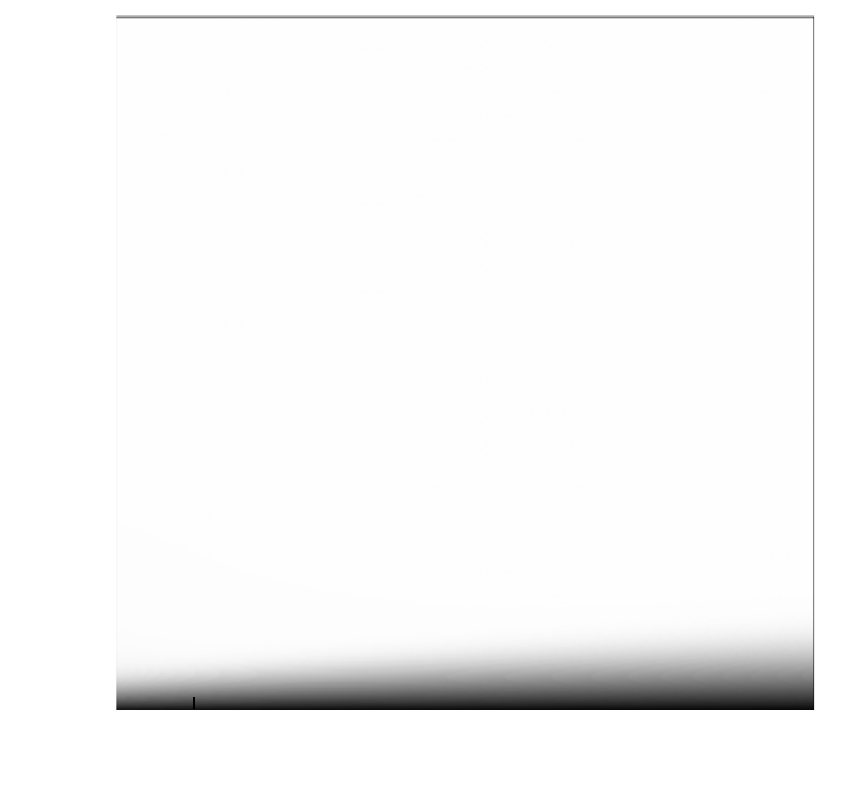
    \end{minipage}\hfill 
    \begin{minipage}{0.48\textwidth}
        \centering
        \def\svgwidth{\linewidth}
        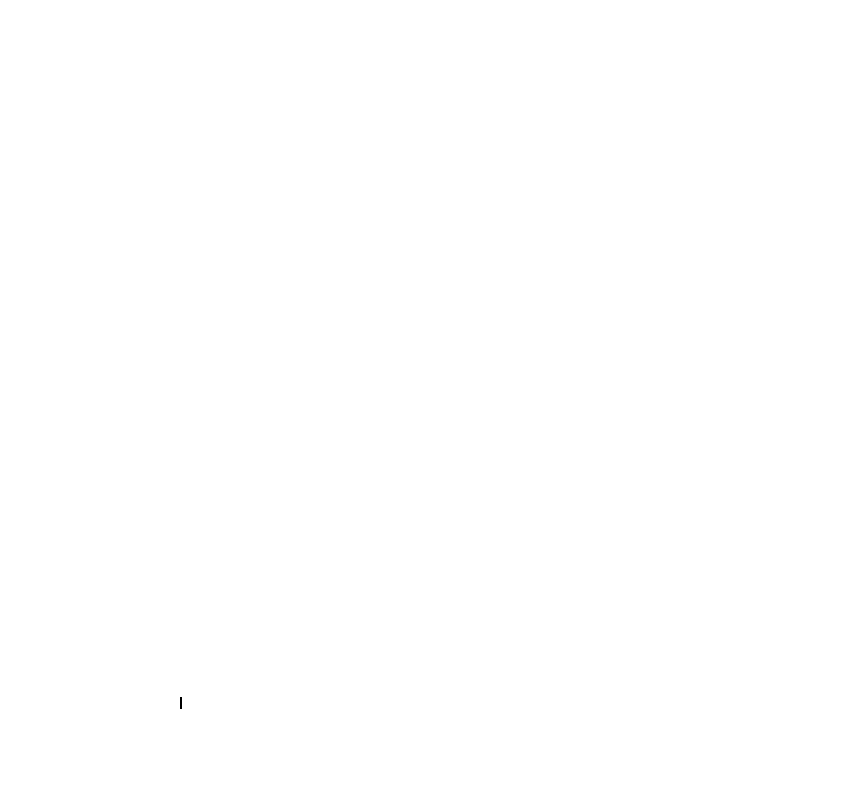
    \end{minipage}
    
    \caption{Grazing study. (\textit{a}) Laminar baseflow and particle trajectories, and (\textit{b}) Hilbert maximum of particle-induced wavepackets.}
    \label{fig:grazing_study}
\end{figure}

The trajectories corresponding to the particle densities listed in table \ref{tab:grazing_study} are shown in figure \ref{fig:grazing_study}(a). The intended design of the study is evident: the particle with $\rho_{p}^{*}=0.50$ g/cm$^{3}$ impinges on the cone surface and reflects, the  particle with $\rho_{p}^{*}=0.25$ g/cm$^{3}$ penetrates deeply into the boundary layer, and the particle with $\rho_{p}^{*}=0.20$ g/cm$^{3}$ 
reaches the middle of the  boundary layer. Hereafter, particle densities are normalised by the maximum considered, such that $\rho_{p}=\rho_{p}^{*}/\rho_{p,max}^{*}$. The inset of figure~\ref{fig:grazing_study}(b) compares wall-pressure amplitude spectral densities extracted at $x^*=0.635$ m from two- and three-dimensional NLDE simulations for the $\rho_p=1.0$ case, with the three-dimensional result evaluated at $\beta^*=0$. The close agreement confirms the two- and three-dimensional equivalence discussed in section \ref{sec:nlde_sec}. The remainder of the analysis therefore employs two-dimensional NLDE simulations, which provide physically meaningful spanwise spectral densities at $\beta^*=0$. 

For each case, the envelope of the excited second-mode wavepacket is obtained at each streamwise location from the maximum magnitude of the Hilbert transform of the wall-pressure disturbance, $\tilde{p}_w$. The resulting distributions are shown in figure~\ref{fig:grazing_study}(b). Despite their markedly different trajectories, the $\rho_p=1.0$ impinging particle and the $\rho_p=0.5$ grazing particle (deep inside the boundary layer) generate wavepackets of nearly identical amplitude downstream. For the shallower grazing trajectory, $\rho_p=0.4$, the wavepacket amplitude increases more gradually as the particle enters the boundary layer, but ultimately reaches a comparable second-mode amplitude. An additional calculation for the $\rho_p=1.0$ case was performed with the nonlinear terms suppressed. The linear and nonlinear solutions are indistinguishable until the onset of saturation near $x^*=0.9$ m, confirming that the disturbance evolution remains linear over the region of interest.

\subsection{Comparison of NLDE and MMS for particle receptivity}

\begin{figure}
    \centering
    \def\svgwidth{0.95\textwidth}
    
    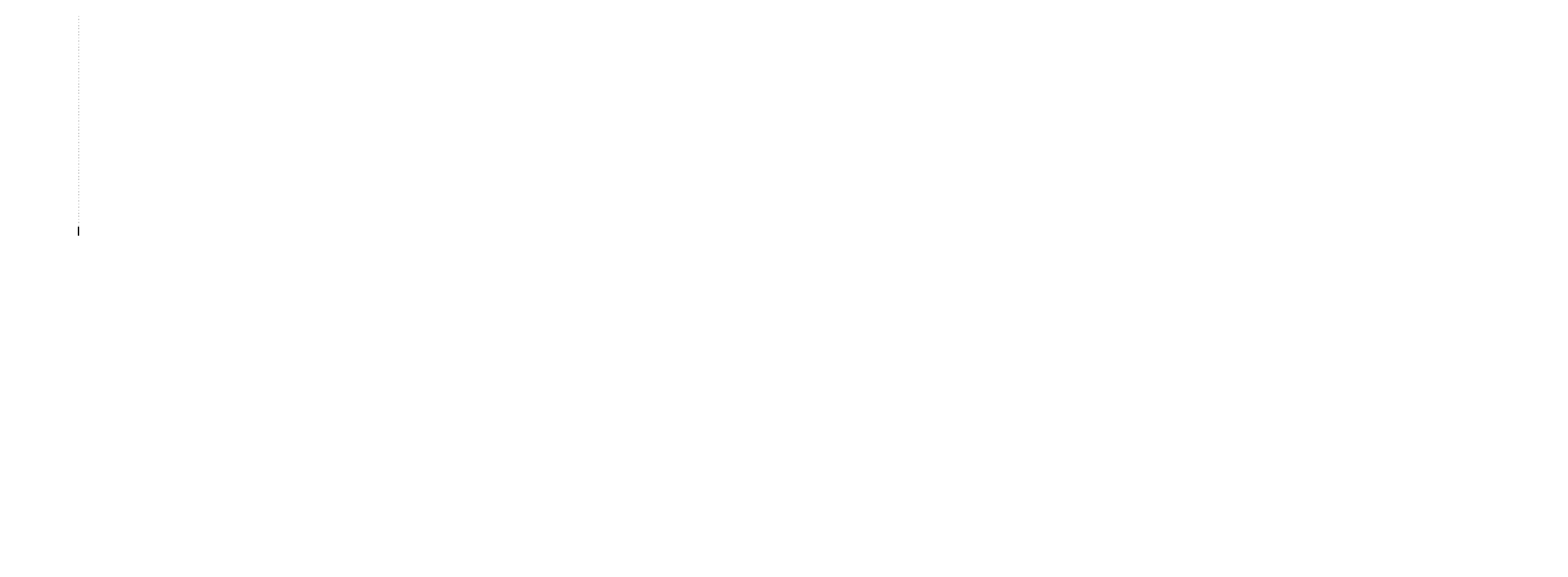
    
    \caption{Comparison of NLDE and MMS particle model.}
    \label{fig:nlde_mosd_comparison_1}
\end{figure}

We now compare the full NLDE simulation with predictions from the MMS particle-receptivity model combined with method of steepest descent denoted as `MMS', for second-mode excitation by  impinging and grazing particles. Figure \ref{fig:nlde_mosd_comparison_1} (top) compares the dimensionless $ASD$ spectra at three streamwise locations. The MMS receptivity model tends to underpredict the spectral amplitudes at frequencies above the dominant frequency, $f_{max}^{*}$. This figure includes a comparison with the particle-receptivity model of \citet{Fedorov2013}, denoted `FM'. This model is evaluated only at $f^{*}=433.7$ kHz for which the neutral point coincides with the particle impingement location and where the temporal formulation is strictly valid, i.e. Gaster transformation. The FM model is not applied to the grazing trajectories because their particle-boundary-layer interaction extends over a finite streamwise distance that cannot be represented accurately within a localised formulation. The bottom row of figure \ref{fig:nlde_mosd_comparison_1} compares the  wall-pressure amplitude, $\tilde{p}_{w}$, normalised by twice dynamic pressure. The MMS prediction is evaluated
by substituting the initial disturbance amplitude spectral densities ($ASD$) computed from equation \ref{eq:mms_asd_eq} into the 2-D steepest-descent expression in equation \ref{eq:2d_3d_mosd}. For all three particle trajectories, MMS accurately predicts the downstream wavepacket amplitude beyond the receptivity region.

\subsection{The receptivity mechanics of impinging and grazing particles}

\begin{figure}
    \centering
    \def\svgwidth{0.95\textwidth}
    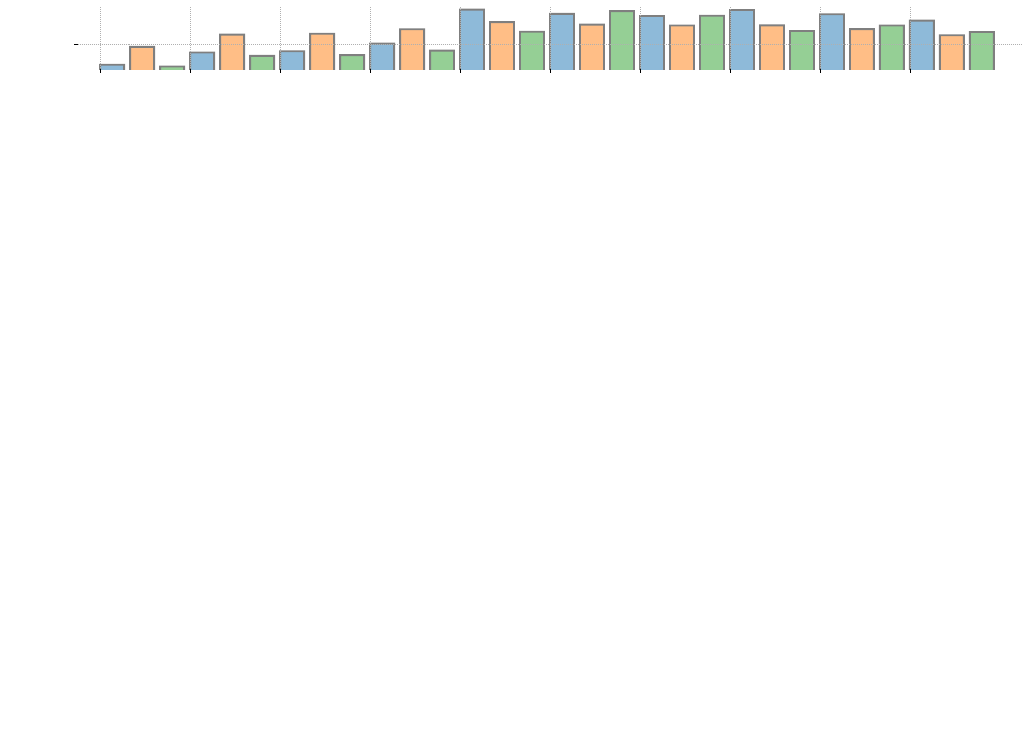
    
    \caption{Forcing distributions for $\rho_{p}=1.0$ (top) and $0.5$ (bottom) at $f^{*}=452$ kHz.}
    \label{fig:forcing_comparison_both}
\end{figure}

We next examine the principle mechanisms governing particle receptivity for impinging and grazing trajectories. The particle forcing term $q$, which drives the amplitude factor $c(x_{1})$ in eq. \ref{eq:amp_ode}, contains three non-zero contributions associated with the streamwise-momentum ($q_{2}$), wall-normal-momentum ($q_{4}$), and energy ($q_{6}$) source terms. Figure \ref{fig:forcing_comparison_both} shows the corresponding forcing distributions prior to wall-normal integration, together with their associated adjoint eigenfunctions (EF), at several streamwise locations for the $\rho_{p}=1.0$ and $\rho_{p}=0.5$ at $f^{*}=452$ kHz (chosen given its high $ASD$ downstream). Each adjoint EF is normalised by its maximum magnitude. Within each column, the bilinear-product distributions, $(G_{i},B_{0,i})$, are normalised by the maximum among the three rows to identify the locally dominant forcing contribution.  The bars in the top row show the maxima of the unaltered $(G_{i},B_{0,i})/\Gamma$ distributions at each streamwise location and therefore provide a global comparison of their relative magnitudes. When the particle is above the boundary layer, indicated by purple dashed-dot line, the wall-normal-momentum component is locally dominant (e.g., $R=2095$). As the particle enters the boundary layer near $R=2200$, the streamwise relative velocity, $u_{rel}=(u_{p}-u_{x})$, increases from approximately zero. At the same time, the particle crosses the critical layer, indicated by pink dashed line, where the adjoint EF components $B_{0,2}$, $B_{0,4}$, and $B_{0,6}$, attain large values. Because $u_{rel}<1$, the streamwise-momentum forcing, which scales linearly with $u_{rel}$, exceeds the energy contribution containing $u_{rel}^2$ and, thus, becomes the dominant local contribution. As the particle approaches the wall at $R=2247$, $u_{rel}$ becomes $\mathcal{O}(1)$ while the energy adjoint EF component $B_{0,6}$ exhibits a second peak. Their combined effect causes the energy forcing to become locally dominant. The maxima shown in the top row of figure \ref{fig:forcing_comparison_both} further demonstrate that the wall-normal-momentum contribution remains relatively small to the streamwise-momentum and energy contributions once the particle has penetrated the boundary layer.

A similar receptivity mechanism is observed for the grazing $\rho_{p}=0.5$ trajectory, shown in the lower half of figure \ref{fig:forcing_comparison_both}. The principal difference is the absence of a substantial energy contribution to $q$, because the particle cannot penetrate deep enough into the boundary layer to attain a sufficiently large relative velocity. Nevertheless, the wall-normal-momentum forcing outside and the streamwise-momentum forcing generated at the critical layer remain active and provide an effective mechanism for disturbance excitation.

\begin{figure}
    \centering
    \def\svgwidth{0.95\textwidth}
    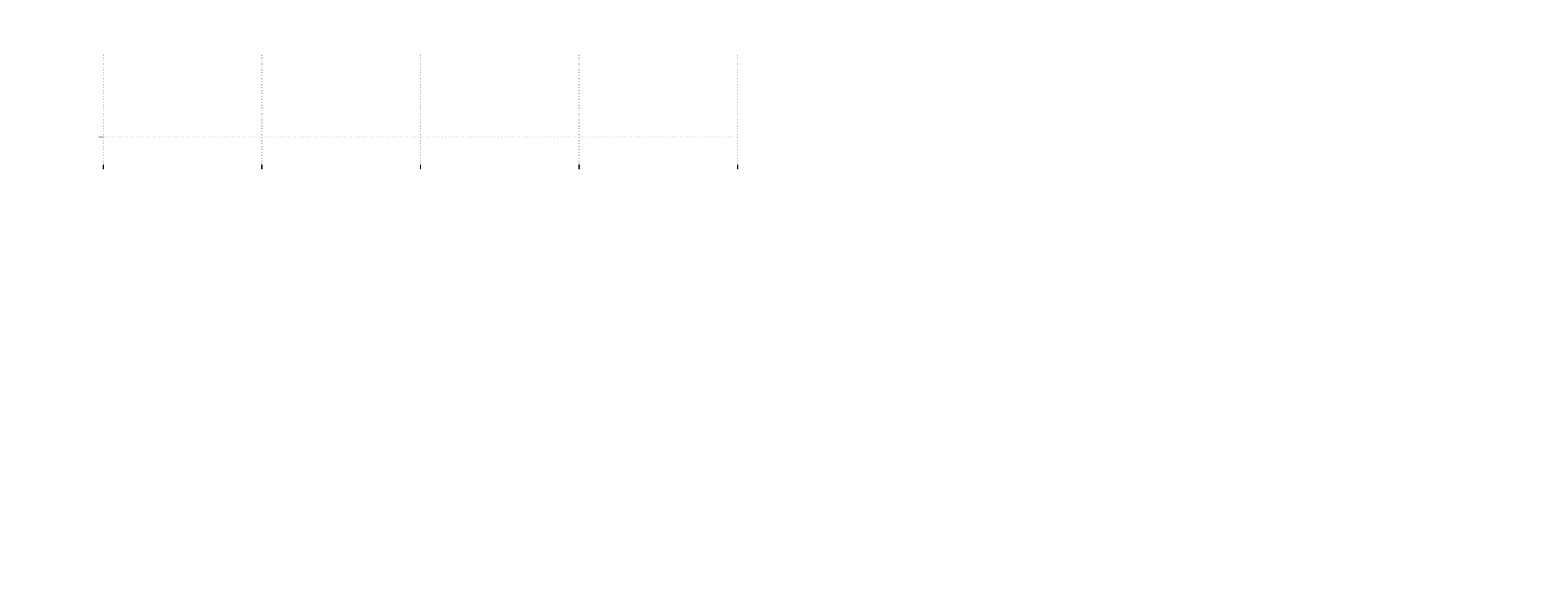
    
    \caption{Components of forcing at $f^*=452$ kHz.}
    \label{fig:forcing_comparison_3}
\end{figure}

Having established that the relative importance of  particle-forcing components is dictated essentially by the particle's position relative to corresponding adjoint EF, we now examine how they combine to complete the receptivity process. The individual components of $q$ in equation \ref{eq:amp_ode}, obtained by integrating the profiles of $(G_{i},B_{0,i})/\Gamma$ in figure \ref{fig:forcing_comparison_both}, are plotted in figure \ref{fig:forcing_comparison_3}, again at $f^{*}=452$ kHz for both trajectory scenarios. For the impinging $\rho_{p}=1.0$ case, the dominant contributions to $q$ arise successively from $q_{4}$ outside the boundary layer, $q_{2}$ as the particle crosses the critical layer, and $q_{6}$ near the wall. The corresponding distributions are roughly symmetric about the impingement location. Each of these forcing regions produces a pronounced increase in the amplitude factor $c(x_{1})$, indicating that the receptivity to an impinging particle proceeds through three distinct stages.

For the grazing $\rho_{p}=0.5$ trajectory, shown on the right-hand side of figure \ref{fig:forcing_comparison_3}, the dominant contributions to $q$ arise from $q_{4}$ outside the boundary layer and $q_{2}$ inside the boundary layer. As noted previously, the grazing particle does not penetrate sufficiently close to the wall for $q_6$ to make significant contributions, and therefore, the receptivity of grazing particles can be interpreted as a two-stage process. Only the qualitative behaviour of $q$ and $c(x_{1})$ is compared between the two trajectories because their magnitudes depend on the upstream integration limit, $x_{1,u}$, which differs between the two cases. The reconstructed amplitude $A(x_{1})$, however, is independent of the choice of $x_{1,u}$, and may therefore be compared directly \citep{Edwards2019}. A key feature of the distributed receptivity process in both cases is the downstream plateau of $c(x_{1})$. This behaviour arises because the factor $K^{-1}$ (grey dash-dot line in figure \ref{fig:forcing_comparison_3})
weighs the forcing contribution most strongly near the neutral point and decays away from it in both streamwise directions. Once the particle has moved sufficiently far downstream of this receptivity region, its forcing is overwhelmed by exponential growth.

\section{Discussion} \label{sec:discussion}
Having developed a theoretical framework capable of predicting both the receptivity and downstream amplification of wavepackets generated by impinging and grazing particles, we now assess the extent to which each trajectory class may contribute to laminar–turbulent transition in hypersonic free flight. Equation 2.15 of \citet{Chuvakhov2025} is used to estimate the root-mean-square (RMS) wall pressure  amplitude of second mode waves excited by particles of a prescribed radius. To extend the formulation to grazing particles, a discrete set of admissible grazing trajectories is searched at each saddle point, $(\omega_s,\beta_s)$, and the trajectory producing the maximum growth is selected. Transition onset is then estimated by applying a threshold criterion to the resulting wall-pressure RMS. 
Predictions are presented in table \ref{tab:rms_comparison} 
for particles with radii $R_{p}^{*}=2$ $\mu$m (all trajectories impinge the cone wall) and $R_{p}^{*}=0.275$ $\mu$m (no trajectories impinge the cone wall). The transition-onset location inferred from the flight measurements of \citet{Borg2025} is also included for comparison.
\begin{table}
  \begin{center}
  \begin{tabular}{cccc}
     & Impinging ($R_{p}^{*}=2$ $\mu$m) & Grazing ($R_{p}^{*}=0.275$ $\mu$m) & Flight \\
    \hline
    $x_{tr}^{*}$ (m) & 0.71 & 1.15 & 1.29 \\
  \end{tabular}
  \caption{Comparison of transition onset locations predicted by impinging and grazing particles 26.95 seconds into flight of HIFLIER vehicle \citep{Borg2025}.}
  \label{tab:rms_comparison}
  \end{center}
\end{table}
The transition-onset location predicted for grazing particles, defined as the position at which the wall-pressure RMS reaches 10\% of the mean boundary-layer-edge pressure, agrees more closely with that inferred from the flight measurements at this stage of HIFLIER's ascent. Moreover, the freestream concentration of submicron particles can exceed that of particles with $R_p=2$ $\mu$m by two orders of magnitude  \citep{habeck2020}. Although each submicron grazing particle generates a lower-amplitude wavepacket, their substantially greater abundance suggests that they were more likely than larger impinging particles to sustain the transition front during this portion of the HIFLIER hypersonic flight.

\section{Conclusion}
This study sought to clarify the role that submicron atmospheric particles following grazing trajectories may play in initiating LTT during hypersonic free flight. A parametric investigation was conducted of Mack's second mode wavepackets excited by solid, spherical particles within the boundary layer of a $7^{\circ}$ straight cone flying Mach 6 at an altitude of 21.5 km, corresponding to the HIFLIER vehicle at 26.95 s into its ascent. Grazing particles have previously been expected to produce substantially weaker disturbances than larger particles that traverse the boundary layer and impinge on the wall \citep{Chuvakhov2025}. The present NLDE simulations demonstrate, however, that the downstream wavepacket amplitudes generated by the grazing trajectories considered are at most 45\% lower than those produced by impinging particles, even though the grazing particles penetrate only partially through the boundary layer before the receptivity process is essentially complete.

To explain this behaviour, a spatial formulation of the particle-receptivity problem was developed using the method of multiple scales. When combined with the method of steepest descent, the formulation accurately reproduces the downstream wavepacket amplitudes obtained from the NLDE simulations. The principal difference between the impinging three-stage and grazing two-stage receptivity mechanisms is the absence of a substantial energy-forcing contribution near the wall. Nevertheless, both trajectories produce strong streamwise-momentum forcing as the particle crosses the critical layer, where the boundary layer is most receptive. This critical-layer interaction accounts for the comparable disturbance amplitudes generated by the two trajectory classes. Finally, the statistical model of \citet{Chuvakhov2025} was used to estimate the RMS wall-pressure amplitude of particle-generated second-mode disturbances. The transition-onset location predicted for grazing particles agrees more closely with that inferred from the HIFLIER flight measurements. Moreover, submicron particles may occur at freestream concentrations up to two orders of magnitude greater than larger particles with $R_p^*=2~\mu\mathrm{m}$ \citep{habeck2020}. Although an individual submicron grazing particle generates a lower-amplitude wavepacket, their substantially greater abundance, together with the strong receptivity associated with critical-layer crossing, suggests that grazing particles may have played a leading role in sustaining the transition front during this stage of the HIFLIER ascent.

\backsection[Acknowledgements]{The authors would like to thank Dr. Kevin Luna for helpful discussions.}

\backsection[Funding]{Sean Dungan was supported by the DoD through the NDSEG Fellowship Program.}

\backsection[Declaration of interests]{The authors report no conflict of interest.}



\bibliographystyle{jfm}
\bibliography{jfm}

\end{document}